\documentclass[english,prl, reprint, groupedaddress, secnumarabic, showkeys, showpacs, raggedbottom]{revtex4-1}
\usepackage[T1]{fontenc}
\usepackage[latin9]{inputenc}
\usepackage{color}
\usepackage{array}
\usepackage{float}
\usepackage{units}
\usepackage{amsmath}
\usepackage{amssymb}
\usepackage{graphicx}

\makeatletter

\providecommand{\tabularnewline}{\\}

\usepackage{natbib,bm,graphicx,url,times,layouts}
\usepackage{txfonts}   
\definecolor{blueviolet}{rgb}{0.2, 0.2, 0.6}
\usepackage[pdftex,          
    bookmarks=false,          
    colorlinks=true,
    allcolors=blueviolet,
    pdfstartview={FitH},  
    ]{hyperref}
\usepackage{pgffor,pdfpages}  
\def\@fnsymbol#1{\ensuremath{\ifcase#1 
  \or \varheartsuit
  \or \clubsuit
  \or \spadesuit
  \or *
  \or \dagger
  \or \ddagger
  \or \mathsection
  \or \mathparagraph
  \or \|
  \or ** 
  \or \dagger\dagger   
  \or \ddagger\ddagger 
  \else\@ctrerr\fi
}}

\makeatother

\usepackage{babel}
\begin{document}
\global\long\def\bra{\langle}
\global\long\def\ket{\rangle}
\global\long\def\half{\frac{1}{2}}
\global\long\def\third{\frac{1}{3}}
\global\long\def\dx{\frac{\partial}{\partial x}}
\global\long\def\dxd{\frac{\partial}{\partial\dot{x}}}
\global\long\def\thrha{\frac{3}{2}}
\global\long\def\sq{\sqrt{2}}
\global\long\def\sqinv{\frac{1}{\sqrt{2}}}
\global\long\def\up{\uparrow}
\global\long\def\do{\downarrow}
\global\long\def\p{\partial}
\global\long\def\dqi{\frac{\partial}{\partial q_{i}}}
\global\long\def\dqid{\frac{\partial}{\partial\dot{q}_{i}}}
\global\long\def\a{\alpha}
\global\long\def\b{\beta}
\global\long\def\g{\gamma}
\global\long\def\c{\chi}
\global\long\def\d{\delta}
\global\long\def\o{\omega}
\global\long\def\m{\mu}
\global\long\def\n{\nu}
\global\long\def\z{\zeta}
\global\long\def\l{\lambda}
\global\long\def\e{\epsilon}
\global\long\def\x{\chi}
\global\long\def\r{\rho}
\global\long\def\t{\theta}
\global\long\def\G{\Gamma}
\global\long\def\D{\Delta}
\global\long\def\O{\Omega}
\global\long\def\L{\Lambda}
\global\long\def\P{\Phi}
\global\long\def\T{\Theta}
\global\long\def\db{\dbar}
\global\long\def\dg{\dagger}
\global\long\def\s{\sigma}
\global\long\def\k{\kappa}
\global\long\def\pb{\pmb{\phi}}
\global\long\def\pbd{\dot{\pmb{\phi}}}
\global\long\def\pbdd{\ddot{\pmb{\phi}}}
\global\long\def\pp{\pmb{\zeta}}
\global\long\def\ppdd{\ddot{\pmb{\zeta}}}
\global\long\def\cap{\mathcal{C}}
\global\long\def\omeg{\mathcal{C}^{-1}\mathcal{L}^{-1}}
\global\long\def\ind{\mathcal{L}^{-1}}
\global\long\def\cn{C_{0}}
\global\long\def\vx{V_{x}}
\global\long\def\vy{V_{y}}
\global\long\def\cnt{\tilde{C}_{0}}
\global\long\def\vxt{\tilde{V}_{x}}
\global\long\def\vyt{\tilde{V}_{y}}
\global\long\def\vxtd{\tilde{V}_{x}^{\dg}}
\global\long\def\vytd{\tilde{V}_{y}(m)^{\dg}}
\global\long\def\R{\tilde{R}}
\global\long\def\OO{\tilde{\O}^{2}}
\global\long\def\uy{U_{y}}
\global\long\def\linv{\ell_{\text{inv}}}
\global\long\def\ta{t_{A}}
\global\long\def\tha{\t_{A}}
\global\long\def\la{\l_{A}}
\global\long\def\tna{t_{NA}}
\global\long\def\thna{\t_{NA}}
\global\long\def\lna{\l_{NA}}
\global\long\def\pin{\pmb{\phi}_{\text{in}}}
\global\long\def\pout{\pmb{\phi}_{\text{out}}}
\global\long\def\sx{\s_{1}}
\global\long\def\sy{\s_{2}}
\global\long\def\sz{\s_{3}}
\global\long\def\S{\Sigma}
\newcommand{\phottop}{raghu2008,*wang2009,*koch2010,*umucalilar2011,*kraus2012b,*fang2012,*ochiai2012,*liangchong,*rechtsman2012,*rechtsman2013,*verbin2013,*lu2013,*davoyan2013,*peano2014,*nalitov2014,*wang2014a,*karzig2014,*bardyn2014}\newcommand{\opttop}{gerbier2013,*aidelsburger2014}\newcommand{\optres}{hafezi2011,*mittal2014}\newcommand{\trihof}{cocks2012,*orth2013,*wang2014}\newcommand{\noncir}{kapit2013,*hafezi2014}\definecolor{cadmiumorange}{RGB}{247, 139, 11}\definecolor{red}{RGB}{255, 0, 0}\definecolor{blue}{RGB}{0, 0, 255}\definecolor{cadmiumgreen}{RGB}{0, 128, 0}\definecolor{gray}{RGB}{205, 197, 197}
\newenvironment{lyxgreyedout}{\textcolor{red}\bgroup}{\egroup}

\preprint{preprint}

\title{Topological properties of linear circuit lattices}

\author{Victor V. Albert}

\author{Leonid I. Glazman}

\author{Liang Jiang}

\affiliation{Departments of Applied Physics and Physics, Yale University, New
Haven, Connecticut, USA}


\date{\today}
\begin{abstract}
Motivated by the topologically insulating (TI) circuit of capacitors
and inductors proposed and tested in arXiv:1309.0878, we present a
related circuit with less elements per site. The normal mode frequency
matrix of our circuit is unitarily equivalent to the hopping matrix
of a quantum spin Hall insulator (QSHI) and we identify perturbations
that do not backscatter the circuit's edge modes. The idea behind
these models is generalized, providing a platform to simulate tunable
and locally accessible lattices with arbitrary complex spin-dependent
hopping of any range. A simulation of a non-Abelian Aharonov-Bohm
effect using such linear circuit designs is discussed.
\end{abstract}
\maketitle
The realization that electrons propagating on edges of two-dimensional
topological insulators at zero temperature are protected from certain
disorder \cite{kane2005,kane2005a,sheng2005,sheng2006,bernevig2006}
has spurred research simulating these and similar edge effects in
photonic/phononic systems \cite{\phottop,\optres,khanikaev2013,he2014}
(reviewed in \cite{lu2014}). The existence of edge modes whose energies
lie within a given bulk gap of a noninteracting tight-binding Hamiltonian
can be traced to a certain property of the corresponding hopping matrix
\cite{kitaev_clas}. Namely, a \textit{topologically nontrivial }hopping
matrix is characterized by having a nontrivial value of some topological
invariant at that bulk gap. Therefore, the problem of engineering
edge modes in bosonic systems can be reduced to making sure that time
evolution is governed by some topologically nontrivial matrix. Many
efforts emulate the electronic systems that inspired us, but over
time we should be able to construct a wider variety of systems than
those readily available in nature (e.g. \cite{stanescu2010}). While
edge mode protection in topologically nontrivial bosonic systems may
not be as intrinsic or robust (e.g. protection is not guaranteed by
time-reversal symmetry; see Box 2 of \cite{lu2014}), these directions
should nevertheless advance understanding and could offer novel applications
of the materials in question.

In this Letter, we discuss topologically insulating (TI) circuits
\cite{jia2013} -- lattices of inductors and capacitors whose normal
mode frequency matrix $\O^{2}$ mimics a topologically nontrivial
hopping matrix of an electronic system. Topological photonics includes
many proposals \cite{\phottop,\optres}; here we study only inductors
and capacitors with the goal of providing the simplest building blocks
that can lead to topological nontriviality. We discuss a minimal example
whose $\O^{2}$ is (unitarily) equivalent to the hopping matrix of
a spinful 2D electron gas in a magnetic field (see Sec. 5.2 in \cite{topobook}),
i.e., a spin-doubled Azbel-Hofstadter model \cite{azbel,*hofstadter1976}
(deemed the time-reversal invariant \textit{(TRI) Hofstadter model}
\cite{\trihof}). Our example simulates $\nicefrac{1}{3}$ magnetic
flux per plaquette. Such a model is (topologically) similar to the
spin-doubled Haldane model lattice \cite{haldane1988} (see Sec. 9.1.2
in \cite{topobook}) that is featured in the more general Kane-Mele
$\mathbb{Z}_{2}$ topological insulator \cite{kane2005,kane2005a}.
We determine how features of such models carry over to the circuit
context, summarized in a table at the end of the article. The first
TI circuit, which has already been realized \cite{jia2013}, is a
simple extension of our example and we outline that design in \footnotemark[1]\footnotetext{See Supplemental Material [URL], which includes Refs. \cite{kibler2009,magma}, for a comparison of this work to \cite{jia2013} as well as details on the non-Abelian generalization.}.
We further generalize the recipe and provide a method to construct
$\O^{2}$ equivalent to the hopping matrix of a lattice of spins with
arbitrary complex spin-dependent hopping. Notably, we show how to
simulate \textit{any} $U(1)$ hopping with a smaller circuit than
that of \cite{jia2013}, which simulated a \textit{specific} $U(1)$
hopping. This provides a platform to synthesize background gauge fields
using linear circuits in parallel to studies with more complex elements
\cite{\noncir,\optres} and to intense investigations using ultracold
atoms (e.g. \cite{\opttop,osterloh2005,jacob2008,bermudez2010} and
refs. therein).

\begin{figure}[H]
\centering{}\includegraphics[width=1\columnwidth]{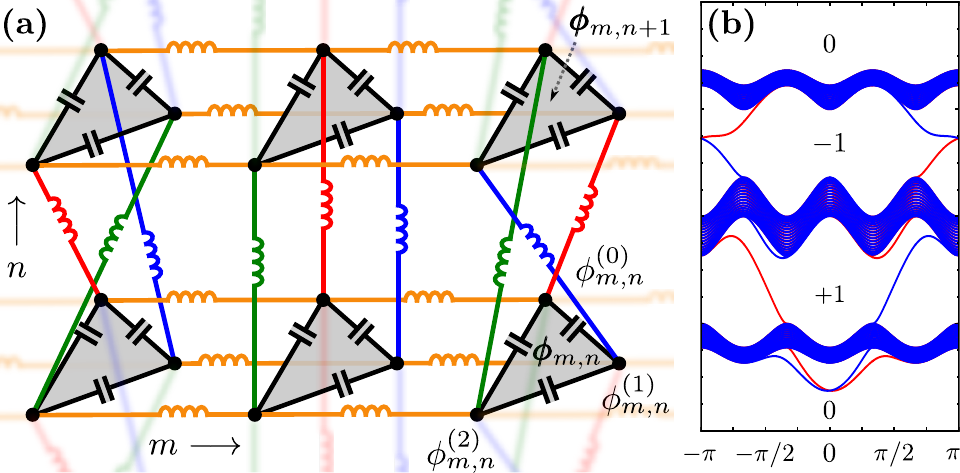}\protect\caption{\label{f1}(color online) \textbf{(a)} Circuit diagram of a TI circuit
lattice, whose normal mode frequency matrix $\protect\O^{2}$ is equivalent
to the hopping matrix of the spin-doubled Hofstadter model in the
Landau gauge with respective $\pm\nicefrac{1}{3}$ magnetic flux per
plaquette. All inductors (capacitors) have uniform inductance (capacitance),
so colors are used for visual aid only. The lattice consists of triangular
sites $m,n$ (labeled as $\protect\pb_{m,n}$, shaded grey), each
consisting of three integrated voltages $\phi_{m,n}^{(\protect\m)}$
($\protect\m=0,1,2$) at its nodes. The vertical inductive connection
is dependent on the horizontal index $m$ and generated by the cyclic
wiring permutation $\protect\vy$ in Eq. (\ref{eq:vy}). \textbf{(b)}
Band structure of $\protect\O^{2}$ simulating a semi-infinite sample,
i.e., a wide vertical strip with the left edge consisting of $(\protect\vy)^{0}$
permutations and right edge mode bands removed. Bands for the spin
up (down) component of the TRI Hofstadter model are in red (blue).
The spin Chern number $\mathcal{C}_{\text{sc}}$ (see text) is written
inside each gap. The edge modes below the lowest bulk band arise because
of circuit edge effects \protect\footnotemark[2] and are not topologically
protected because they do not traverse a gap.}
\end{figure}

\footnotetext{Circuit edge effects distort the original TRI Hofstadter spectrum: the 4 from Eq.~(\ref{eq:two}) is replaced by a 3 (2) for sites on edges (corners). Edge modes exist for all three different types of edges of a vertical strip.}

\textit{Minimal example.}---We distill the idea from \cite{jia2013}
in the form of a simplified example (Fig. \ref{f1}a) and detail how
our methods and conclusions apply to \cite{jia2013} elsewhere \footnotemark[1].
Our circuit consists of a lattice of \textit{sites} (gray), each site
consisting of three \textit{nodes}. Inductors link sites to each other
while capacitors couple the nodes within a site. We stress that no
external flux is threaded through any loop of the circuit and the
magnetic flux of the Hofstadter model is \textit{simulated} via the
intersite inductive wiring. Transforming the real normal mode frequency
matrix $\O^{2}$ into the form of a Hofstadter hopping matrix consists
of grouping degrees of freedom into vectors and performing a transformation
to complex variables. In an ungrounded circuit, each node $m,n,\m$
(with $\m=0,1,2$ labeling the degrees of freedom of the site) has
a time-integrated absolute voltage $\phi_{m,n}^{(\m)}\equiv\int_{-\infty}^{t}v_{m,n}^{(\m)}(t^{\prime})dt^{\prime}$
associated with it \cite{devoret1995}. This labeling scheme introduces
redundant degrees of freedom (which will soon be removed), but allows
$\O^{2}$ to be determined analytically. We now group the nodes at
each site $m,n$ into a \textit{vector} $\pb_{m,n}^{\mathsf{T}}=\bra\phi_{m,n}^{(0)},\phi_{m,n}^{(1)},\phi_{m,n}^{(2)}\ket$.
For example, the Lagrangian contribution of the link between site
$m,n$ and $m,n+1$ (see Fig. \ref{f1}a) is then organized into a
(kinetic) capacitive part $\half\sum_{\d=0,1}\pbd_{m,n+\d}^{\mathsf{T}}C_{0}\pbd_{m,n+\d}$
and a (potential) inductive part 
\[
{\textstyle \half}(\sum_{\d=0,1}\pb_{m,n+\d}^{\mathsf{T}}I_{3}\pb_{m,n+\d}-\pb_{m,n}^{\mathsf{T}}\vy\pb_{m,n+1}-\pb_{m,n+1}^{\mathsf{T}}\vy^{\mathsf{T}}\pb_{m,n})
\]
with $I_{n}$ $n\times n$ identity and respective onsite/intersite
couplings
\begin{equation}
\cn=\frac{1}{3}\begin{pmatrix}2 & -1 & -1\\
-1 & 2 & -1\\
-1 & -1 & 2
\end{pmatrix}\,\,\,\,\,\,\,\,\,\,\text{and}\,\,\,\,\,\,\,\,\,\,\vy=\begin{pmatrix}0 & {\color{red}1} & 0\\
0 & 0 & {\color{blue}1}\\
{\color{cadmiumgreen}1} & 0 & 0
\end{pmatrix}\,.\label{eq:vy}
\end{equation}
Above, the colored matrix elements correspond respectively to the
${\color{red}\text{red}}$, ${\color{blue}\text{blue}}$, and ${\color{cadmiumgreen}\text{green}}$
circuit elements from Fig. \ref{f1}a, and we have set a uniform capacitance
of a third (for normalization) and inductance of one. The equation
of motion (EOM) for $\pb_{m,n}$ in the lattice from Fig. \ref{f1}a
is
\begin{eqnarray}
\cn\pbdd_{m,n} & = & -4\pb_{m,n}+V_{x}\pb_{m+1,n}+V_{x}^{\mathsf{T}}\pb_{m-1,n}\label{eq:1}\\
 &  & \,\,\,\,\,\,\,\,\,\,+(\vy)^{m}\pb_{m,n+1}+(\vy^{\mathsf{T}})^{m}\pb_{m,n-1}\,,\nonumber 
\end{eqnarray}
where $\vx=I_{3}$ and 4 is the number of nearest neighbors for a
site in the bulk. The three distinct powers of $\vy$ {[}$(\vy)^{3}=I_{3}${]}
correspond to three vertical inductive wiring permutations and mimic
the Hofstadter model in the Landau gauge.

To diagonalize $\O^{2}$ in the index $\m$ and simultaneously remove
the aforementioned redundant degrees of freedom, one can apply a discrete
Fourier transform $F$ to the three nodes of each site: $\pp_{m,n}=F\pb_{m,n}$
or $\zeta_{m,n}^{(\m)}=\frac{1}{\sqrt{3}}e^{i\frac{2\pi}{3}\m\n}\phi_{m,n}^{(\n)}$
($\m,\n\in\{0,1,2\}$ and repeated indices summed). This site-preserving
transformation to a \textit{complex} vector $\pp_{m,n}^{\mathsf{T}}=\bra\zeta_{m,n}^{(0)},\zeta_{m,n}^{(1)},\zeta_{m,n}^{(2)}\ket$
block-diagonalizes $\O^{2}$ in $\m$ at the expense of introducing
complex numbers. In the $\pp$ basis, the simultaneously diagonal
capacitive and inductive coupling matrices are $\cnt=\text{diag}(0,1,1)$,
$\vyt=\text{diag}(1,e^{i\frac{2\pi}{3}},e^{-i\frac{2\pi}{3}})$, and
$\vxt=V_{x}=I_{3}$. Since the transformed circuit Lagrangian does
not contain $\dot{\zeta}_{m,n}^{(0)}$ terms (since $(\tilde{C}_{0})_{00}=0$),
the $\zeta_{m,n}^{(0)}\equiv\sum_{\m}\phi_{m,n}^{(\m)}$ component
for each site represents ``half'' of a degree of freedom (akin to
a classical harmonic oscillator in the limit of zero mass) and can
be thought of as an ordinary normal mode in the limit of zero capacitance.
The EOM for $\{\zeta_{m,n}^{(1)},\zeta_{m,n}^{(1)\star}=\zeta_{m,n}^{(2)}\}$,
treated as independent full degrees of freedom ($j=1,2$), is
\begin{equation}
\ddot{\zeta}_{m,n}^{(j)}=-4\zeta_{m,n}^{(j)}+\zeta_{m+1,n}^{(j)}+\zeta_{m-1,n}^{(j)}+e^{i\frac{2\pi}{3}mj}\zeta_{m,n+1}^{(j)}+e^{-i\frac{2\pi}{3}mj}\zeta_{m,n-1}^{(j)}\,.\label{eq:two}
\end{equation}
These variables are linear superpositions of bosonic modes and their
hopping properties resemble the TRI Hofstadter model in the Landau
gauge, i.e., they acquire a (simulated) Peierls phase upon a vertical
hopping. Thus, the block-diagonal normal mode frequency matrix $\OO=\bigoplus_{\m}\OO_{\m}$
consists of the trivial mode matrix $\OO_{0}$ and the matrices $\OO_{1,2}$
forming the spin-doubled Hofstadter model. 

\textit{Topological invariant.---}In Fig. \ref{f1}b, the band structure
of $\OO_{1}$ ($\OO_{2}$) is plotted in red (blue), depicting slightly
distorted \footnotemark[2] counterpropagating edge modes. Since the
pseudo-spin $\bra\zeta^{(1)},\zeta^{(2)}\ket$ is conserved, the spin-doubled
Hofstadter model is characterized by the $\mathbb{Z}$ spin Chern
number $\mathcal{C}_{\text{sc}}={\textstyle \half}(\mathcal{C}_{1}-\mathcal{C}_{2})$
\cite{sheng2006} at each gap. Given an edge, the Chern numbers $\mathcal{C}_{j}$
are related to the number of times the edge modes of $\OO_{j}$ wind
around a horizontal line drawn in the gap (Secs. 5.3.1 and 6.4 in
\cite{topobook}). Moreover, the quantity $\mathcal{C}=\mathcal{C}_{\text{sc}}\text{mod}2$
determines whether there is an even or odd number of pairs of counterpropagating
edge modes (this is the invariant of the more general $\mathbb{Z}_{2}$
TI \cite{kane2005a}, a QSHI with no spin conservation). The invariant
$\mathcal{C}$ is characterized by Kramers degeneracy, which prohibits
elastic backscattering between counterpropagating edge modes only
for odd numbers of edge mode pairs per edge \cite{qi2011}. Both our
example and \cite{jia2013} contain one gapless edge mode pair per
edge ($\mathcal{C}_{\text{sc}}=1$) and, since pseudo-spin is conserved,
constitute a QSHI. Moreover, this system is not a crystalline topological
insulator \cite{fu2011} (as defined in \cite{deleeuw2014}) since
$\mathcal{C}\neq0$.

Due to the invariants established above, there must exist some operator
in the circuit context that mimics the antiunitary electronic time-reversal
operator $i\sy K$ (with $Ki=-iK$ and $\s_{1,2,3}$ the usual Pauli
matrices), squares to $-I_{2}$, and generates a Kramers degeneracy
(a similar observation has been made \cite{he2014} with photonic
TIs \cite{khanikaev2013}). Such an operator does indeed exist and
comes about from a symmetry of the circuit. In the $\pb$ basis, the
coupling matrix $\vy$, a cyclic permutation of all nodes in each
site, commutes with $\O^{2}$ and generates the symmetry group $C_{3}\approx\{I_{3},\vy,V_{y}^{\mathsf{T}}\}$.
A generic linear commuting operator (with identity components in the
dimensions indexed by $m,n$) can be expressed as $c_{\m}(\vy)^{\m}$
for some $c_{\m=0,1,2}\in\mathbb{C}$. Since $\vy$ is real, all antilinear
extensions of the above operators can be expressed as $c_{\m}(\vy)^{\m}K$.
In the $\pp$ basis, 
\[
K\rightarrow\tilde{K}=F^{\dg}KF=F^{\dg}F^{\star}K=(1\oplus\sx)K\,,
\]
which squares to $I_{3}$. However, the operator $S$ {[}such that
$\tilde{S}=(1\oplus\sy)K$ and $\tilde{S}^{2}=1\oplus(-I_{2})${]}
is also in the span of $(\vy)^{\m}K$. Thus, electronic time-reversal
symmetry in the tight-binding context maps to a combination of ordinary
time-reversal and cyclic permutations in the circuit context. We also
note that $\tilde{\Sigma}=\tilde{S}\tilde{K}=1\oplus(-i\sz)$ characterizes
the conserved pseudo-spin for the time-reversed Hofstadter copies.

\textit{Symmetry protection}.---Mirroring topological protection in
QSHIs and $\mathbb{Z}_{2}$ TIs, counterpropagating edge modes of
a TI circuit must also be ``protected'' to some degree. Emulating
one-particle elastic scattering processes in TRI electronic systems
\cite{qi2011}, a crossing between edge modes on the same edge at
time-reversal invariant points $k=0,\pi$ in the Brillouin zone will
not be lifted by inductance or capacitance perturbations that commute
with $S$ (which is now in the $\pb$ basis). Let a generic inductive
link between sites $m,n$ and $p,q$ be parametrized by 
\begin{equation}
\pb_{m,n}^{\mathsf{T}}M_{11}\pb_{m,n}+\pb_{p,q}^{\mathsf{T}}M_{22}\pb_{p,q}+\pb_{m,n}^{\mathsf{T}}M_{12}\pb_{p,q}+\pb_{p,q}^{\mathsf{T}}M_{12}^{\mathsf{T}}\pb_{m,n}\,,\label{eq:3}
\end{equation}
where real $3\times3$ matrices $M_{jj}$ ($j=1,2$) are onsite couplings
at the two respective sites and $M_{12}$ is the intersite coupling.
Such a perturbation will not cause elastic backscattering between
edge modes whenever $[M_{jj^{\prime}},S]=0$. For our design, such
perturbations are all those which do not break the circuit's $C_{3}$
symmetry, i.e., commute with $\vy$. For example, an identical simultaneous
perturbation of all three inductances in any given link {[}$M_{jj}\propto I_{3}$,
$M_{12}\propto(\vy)^{\m}${]} or an onsite perturbation ($M_{jj^{\prime}}\propto\d_{j1}\d_{j^{\prime}1}[(\vy)^{\m}+(\vy^{\mathsf{T}})^{\m}]$)
will not mix edge modes. However, fluctuations of inductance \textit{will}
cause elastic backscattering between edge modes whenever the fluctuations
are \textit{not} identical within any given link. A similar statement
holds for capacitive perturbations.

Topologically insulating circuits (i.e., both our design and \cite{jia2013})
turn out to be similar to certain optical resonator designs \cite{\optres}
in that both are robust against disorder that does not induce flips
of pseudo-spin \cite{lu2014}. In our design, the pseudo-spin is characterized
by $\S=SK$: since $M_{jj^{\prime}}$ are real matrices, $[M_{jj^{\prime}},S]=0\leftrightarrow[M_{jj^{\prime}},\S]=0$.
We also note that, in a realistic setup, both optical resonator edge
states and TI circuit edge modes will decay due to optical and microwave
dissipation, respectively.

\begin{figure}[h]
\includegraphics[width=1\columnwidth]{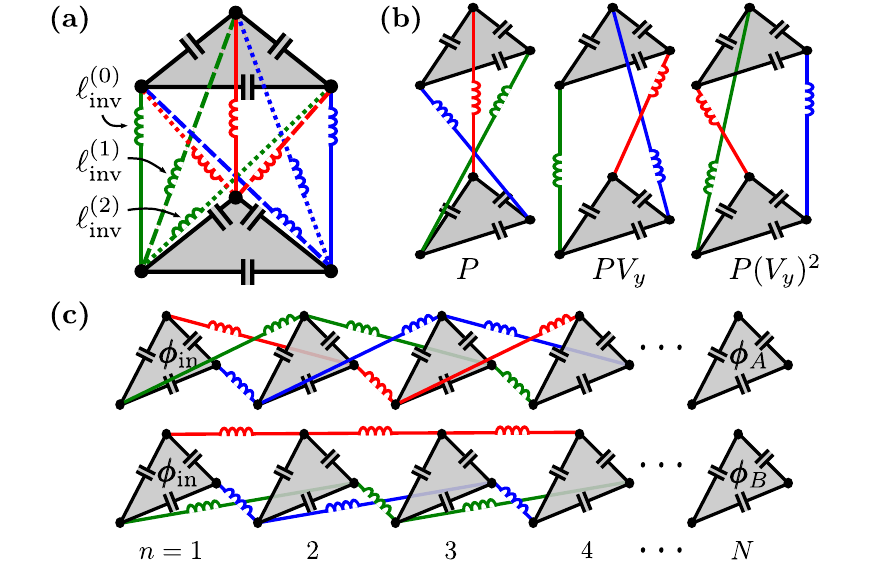}\protect\caption{\label{f2}\textbf{ }(color online) \textbf{(a)} Superposition of
three different wiring permutations $(\protect\vy)^{\protect\m}$
and their respective inverse inductances $\protect\linv^{(\protect\m)}$,
$\protect\m=0,1,2$ (solid, dashed, dotted respectively), achieving
any $U(1)$ hopping in the $\protect\pp$ basis. \textbf{(b)} Additional
wiring permutations $P(\protect\vy)^{\protect\m}$ which create $U(2)$
hopping terms in the $\protect\pp$ basis. \textbf{(c)} A circuit
to simulate the Aharonov-Bohm (AB) effect. A vector signal $\protect\pin$
enters from the left, propagates through $N$ sites via two different
paths $A$ and $B$, and produces two outputs, $\protect\pb_{A,B}$.
One can measure an interference between these outputs {[}Eq. (\ref{eq:ab2}){]}
and observe oscillations for even $N$ since permutations $\protect\vy$
and $P$ do not commute.}
\end{figure}

\textit{Generalizations.}\textbf{---}Given that the above design only
has $d=3$ nodes per site, one can consider increasing the number
of nodes per site (triangles $\rightarrow$ $d$-gons) and generalizing
the cyclic permutation ($\vy\rightarrow\sum_{\m=0}^{d-1}|\m\ket\bra\m+1|\text{\,\ mod\,}d$).
This results in a family of models that can emulate TRI Hofstadter
hopping matrices with $\nicefrac{p}{d}$ background magnetic flux
using $d$ nodes per site and vertical connections $(\vy)^{p}$ (with
integer $p$). We note in passing that the $d=2$ case is trivial
because it is not gapped in the bulk (see Eq. (5.53) in \cite{topobook})
and that \cite{jia2013} is closely related to $d=4$ \footnotemark[1].
However, we have developed other generalizations which allow simulation
of any background gauge field using circuits that are much more compact.
We discuss these approaches below.

First, an \textit{arbitrary} complex hopping can be achieved using
only three nodes per site. For simplicity, we first focus on one link.
Instead of having one wiring permutation (e.g. $\vy$ in Fig. \ref{f1}),
one can implement all three permutations $(\vy)^{\m}$ in a linear
superposition (Fig. \ref{f2}a). In this case, each permutation gains
its own degree of freedom. The intersite inductive coupling matrix
is then $\vy\rightarrow V_{A}=\linv^{(\m)}(\vy)^{\m}$, where $\linv^{(\m)}$
is the inverse inductance of permutation $\m$. In the $\pp$ basis,
the coupling is diagonal with $(\tilde{V}_{A})_{\m\n}=\linv^{(\tau)}e^{i\frac{2\pi}{3}\tau\n}\d_{\m\n}$
(no sum over $\n$). Parameterizing the $\m=1$ component in terms
of an amplitude/phase obtains $(\tilde{V}_{A})_{11}=\ta e^{i\tha}$
with
\begin{eqnarray}
\ta & = & \sqrt{[\linv^{(0)}-{\textstyle \half}(\linv^{(1)}+\linv^{(2)})]^{2}+{\textstyle \frac{3}{4}}(\linv^{(1)}-\linv^{(2)})^{2}}\,,\nonumber \\
\tha & = & \tan^{-1}\left(\frac{\sqrt{3}(\linv^{(1)}-\linv^{(2)})}{2\linv^{(0)}-(\linv^{(1)}+\linv^{(2)})}\right)\,.
\end{eqnarray}
Naturally, $(\tilde{V}_{A})_{00}=\sum_{\m}\linv^{(\m)}\equiv\la$
and $(\tilde{V}_{A})_{22}=\ta e^{-i\tha}$. Additionally, there is
a diagonal inductance contribution of $\half\la\pp^{\dg}\pp$ to both
of the linked sites. Thus, the hopping and diagonal terms $\{\ta,\tha,\la\}$
can be tuned using $\{\linv^{(\m)}\}_{\m=0}^{2}$ with the constraint
$\la\geq\ta$ since $\linv^{(\m)}\geq0$. The symmetry protection
still holds here since $(\vy)^{\m}\in C_{3}$.

Second, non-Abelian couplings can straightforwardly be implemented
while still keeping $d=3$. Instead of using the permutations $(\vy)^{\m}$,
three other permutations $P(\vy)^{\m}$ (with $P=1\oplus\sx$ and
$[P,\vy]\neq0$; see Fig. \ref{f2}b) can be superimposed to give
an inverse inductance coupling matrix $\vy\rightarrow V_{NA}=\linv^{(\m)}P(\vy)^{\m}$.
Nonzero entries of $\tilde{V}_{NA}$ are an off-diagonal hopping $(\tilde{V}_{NA})_{12}=(\tilde{V}_{NA})_{21}^{\star}\equiv\tna e^{i\thna}$
and a diagonal contribution $(\tilde{V}_{NA})_{00}=\sum_{\m}\linv^{(\m)}\equiv\l_{NA}$.
Similar to $V_{A}$, the hopping and diagonal terms $\{t_{NA},\t_{NA},\l_{NA}\}$
of $V_{NA}$ can be tuned using $\{\linv^{(\m)}\}_{\m=0}^{2}$. As
an example, one can already realize a non-Abelian generalization of
the Hofstadter model \cite{osterloh2005} by letting $\vx\rightarrow P$
in Eq. (\ref{eq:1}). 

The above design allows one to create a lattice with spatially nonuniforn
noncommuting unitary hoppings between sites {[}e.g. $t_{m,n}\exp(i\t_{m,n})$
using either $(\vy)^{\m}$ or $P(\vy)^{\m}${]} while maintaining
identical onsite contributions ($\l_{m,n}\equiv\l$). Despite this
flexibility, one cannot create arbitrary $U(2)$ hoppings using three
nodes per site (assuming onsite contributions are to remain identical).
This is because linear superpositions of the six permutations {[}($\vyt)^{\m}$
and $P(\vyt)^{\m}${]} with \textit{nonnegative real coefficients}
(since our variables are inverse inductances) do not span all unitary
$2\times2$ matrices acting on $\bra\zeta^{(1)},\zeta^{(2)}\ket$.
More permutations are needed, so one needs more nodes per site to
generate them. Finding this minimal number of nodes maps to an open
problem from group theory \cite{saunders2008,elias2010}, and we have
determined \footnotemark[1] that one needs at most $n^{2}$ nodes
per site to simulate unitary hoppings of dimension $n>2$.

\textit{Non-Abelian Aharonov-Bohm effect.}\textbf{---}We finish with
a discussion of applications. First we propose an experiment that
uses the $\pb$-$\pp$ duality to observe an electrical non-Abelian
Aharonov-Bohm (AB) effect \cite{fang2013,osterloh2005,jacob2008}.
Since all circuit elements are reciprocal here, it is the non-reciprocity
of their permutations that leads to interference effects. One can
think of $\pb$ as the wavefunctions and sites $n=1,2,...,N$ as spatial
positions (Fig. \ref{f2}c). An incoming signal $\pin^{\mathsf{T}}=\bra\phi_{\text{in}}^{(0)},\phi_{\text{in}}^{(1)},\phi_{\text{in}}^{(2)}\ket$
is applied onto paths $A$ and $B$. Let
\begin{equation}
\phi_{\text{in}}^{(\m)}={\textstyle \sqrt{\frac{2}{3}}\cos(\o t-\frac{2\pi}{3}\m)}\,,\label{eq:ab1}
\end{equation}
which is equivalent to $\pp_{\text{in}}^{\mathsf{T}}=\frac{1}{\sqrt{2}}\bra0,e^{i\o t},e^{-i\o t}\ket$.
Path $A$ contains $N-1$ cyclic permutations $\vy$ from Eq. (\ref{eq:vy})
while path $B$ consists of $N-1$ permutations $P$ from Fig. \ref{f2}b
(with $[\vy,P]\neq0$). Remembering Eq. (\ref{eq:two}), we see that
a phase of $e^{i\frac{2\pi}{3}}$ ($e^{-i\frac{2\pi}{3}})$ is gained
by $\zeta^{(1)}$ ($\zeta^{(2)}$) as the signal ``hops'' sites
in path $A.$ For path $B$, the $\zeta^{(1)}$ and $\zeta^{(2)}$
components are exchanged upon each application of $P$. One can superimpose
the outputs $\pb_{A}$ and $\pb_{B}$ to observe their interference.
For odd $N$, this interference is constant in time. For even $N$,
one should see oscillations due to a nontrivial path $B$:
\begin{equation}
|\pb_{A}+\pb_{B}|^{2}\propto\cos^{2}\{\o t-{\textstyle \frac{2\pi}{3}}\left[\left(N-1\right)\text{mod}3\right]\}\,.\label{eq:ab2}
\end{equation}
Since voltage is the derivative of $\phi$, one can perform the above
experiment by applying voltage signals of the form of $\pb_{\text{in}}$
from Eq. (\ref{eq:ab1}), measuring the six output signals at site
$N$ for paths $A$ and $B$, and superimposing them in the manner
of Eq. (\ref{eq:ab2}). Since the AB effect is nonreciprocal, driving
from right to left ($\pin\leftrightarrow\pb_{A,B})$ should flip the
sign of the phase gained along $A$.

\textit{Outlook.}\textbf{---}This work generalizes the first realization
of a topologically insulating (TI) circuit \cite{jia2013}. We present
a simplified circuit whose normal mode frequency matrix is unitarily
equivalent to the hopping matrix of the time-reversal invariant Hofstadter
model \cite{\trihof} with $\nicefrac{1}{3}$ magnetic flux per plaquette.
A summary of the equivalence is in Table \ref{tab:tc}.
Since Hofstadter models posses edge modes, we determine which perturbations
do not cause edge modes to backscatter.

Additionally, we generalize the approach and determine the minimal
circuit complexity required to simulate non-Abelian background gauge
fields. Besides a simulation of the Aharonov-Bohm effect, we now speculate
on further applications of this circuit QED simulation tool \cite{aspuru-guzik2012,*houck2012,*ashhab2014}.
A major flexibility is being able to construct and locally probe virtually
any lattices (e.g. honeycomb \cite{bermudez2010} or Kagome \cite{petrescu2012})
and lattices with connections other than nearest neighbor at the same
cost in complexity. Almost any physically relevant and exotic geometry
can be implemented \cite{tsomokos2010} (e.g. a Möbius strip \cite{jia2013}).
One can construct interfaces of lattices and observe mixing of edge
modes at the boundary, akin to graphene p-n junctions \cite{abanin2007}.
To simulate interactions, one can substitute Josephson junctions \cite{girvinbook}
(mechanical oscillators \cite{palomaki2013,*braginsky}) for inductors
(capacitors). These and other topics are currently under investigation.

\begin{flushleft}
\begin{table}[t]
\begin{centering}
\begin{tabular}{>{\raggedright}p{100pt}>{\raggedright}p{141pt}}
\hline 
{\small{}TRI Hofstadter model } & {\small{}TI circuit }\tabularnewline
\hline 
{\small{}Hopping matrix} & {\small{}Normal mode frequency matrix $\O^{2}$ }\tabularnewline
{\small{}Fermion $c_{m,n}=\bra c_{m,n}^{(1)},c_{m,n}^{(2)}\ket$} & {\small{}$\zeta_{m,n}=\bra\zeta_{m,n}^{(1)},\zeta_{m,n}^{(1)\star}\ket$
with ${\textstyle \zeta_{m,n}^{(1)}=e^{i\frac{2\pi}{3}\n}\phi_{m,n}^{(\n)}}$}\tabularnewline
{\small{}Peierls phase} & {\small{}Intersite wiring permutations}\tabularnewline
{\small{}Kramers degeneracy} & {\small{}$\tilde{S}=(1\oplus\sy)K$ due to $C_{3}$ symmetry}\tabularnewline[0.04cm]
\hline 
\end{tabular}
\par\end{centering}

\protect\caption{\label{tab:tc}Summary of the equivalence between the Hofstadter model
and a TI circuit. $\phi_{m,n}^{(\mu)}$ is the integrated voltage
at node $m,n,\mu$, as depicted in Fig. \ref{f1}a, $\protect\sy$
is the second Pauli matrix, and $Ki=-iK$.}
\end{table}

\par\end{flushleft}
\begin{acknowledgments}
We thank J. Simon, D. Schuster, A. Dua, T. Morimoto, B. Elias, W.
C. Smith, S. M. Girvin, M. H. Devoret, B. Bradlyn, Z. Minev, and A.
Petrescu for fruitful discussions. We thank one of the referees for
pointing out the possibility of simulating a magnetic flux of $\nicefrac{p}{d}$
with $p\neq1$. This work was supported, in part, by the NSF Graduate
Research Fellowship Program under Grant DGE-1122492 (V. V. A.); NSF
Grant DMR-1206612 (L. I. G.); and the Army Research Office, Air Force
Office of Scientific Research Multidisciplinary Research Program of
the University Research Initiative, Defense Advanced Research Projects
Agency Quiness program, the Alfred P. Sloan Foundation, and the David
and Lucile Packard Foundation (L. J.).
\end{acknowledgments}

\bibliographystyle{apsrev4-1}
\bibliography{C:/Users/Victor/library}



\section*{Supplementary material}

# Supplemental material: Topological properties of linear circuit lattices

Victor V. Albert, Leonid I. Glazman, and Liang Jiang
*Departments of Applied Physics and Physics, Yale University, New Haven, Connecticut, USA*
(Dated: March 10, 2015)

## COMPARISON WITH THE ORIGINAL TI CIRCUIT

The circuit from [1] is similar to the example in the main text, except that the number of nodes per site $d = 4$ instead of 3 and the role of inductances and capacitances is switched. With these differences, $\Omega^2$ will be equivalent to *two* sets of *zero-frequency* modes ($\zeta_{n,m}^{(0)}$ and $\zeta_{m,n}^{(1)}$, akin to classical harmonic oscillators in the limit of zero spring constant) and the *inverse* of the TRI Hofstadter tight-binding matrix with *one-fourth* magnetic flux per plaquette ($\zeta_{n,m}^{(2)}$ and $\zeta_{m,n}^{(3)}$). Instead of $V_y$ generating $C_3$, the $4 \times 4$ intersite capacitive coupling

$$U_y = \begin{pmatrix} 0 & 0 & 0 & 1 \\ 0 & 0 & 1 & 0 \\ 1 & 0 & 0 & 0 \\ 0 & 1 & 0 & 0 \end{pmatrix}$$

generates $C_4$. The coupling $U_y$ is isomorphic to a generalized version of $V_y$ if one cyclically permutes the last three nodes:

$$\begin{pmatrix} 1 & 0 & 0 & 0 \\ 0 & 0 & 0 & 1 \\ 0 & 1 & 0 & 0 \\ 0 & 0 & 1 & 0 \end{pmatrix} \begin{pmatrix} 0 & 0 & 0 & 1 \\ 0 & 0 & 1 & 0 \\ 1 & 0 & 0 & 0 \\ 0 & 1 & 0 & 0 \end{pmatrix} \begin{pmatrix} 1 & 0 & 0 & 0 \\ 0 & 0 & 1 & 0 \\ 0 & 0 & 0 & 1 \\ 0 & 1 & 0 & 0 \end{pmatrix} = \begin{pmatrix} 0 & 1 & 0 & 0 \\ 0 & 0 & 1 & 0 \\ 0 & 0 & 0 & 1 \\ 1 & 0 & 0 & 0 \end{pmatrix}.$$

The antilinear operator $S$ [with $\tilde{S} = (I_2 \oplus \sigma_2)K$ and $\tilde{S}^2 = (I_2 \oplus -I_2)$] lies in the span of $(U_y)^\mu K$ (with $\mu = 0, 1, 2, 3$), revealing a Kramers degeneracy for the conjugate Hofstadter models. Perturbations, of the form of Eq. (4) but this time with $4 \times 4$ matrices $M_{jj'}$, do not cause edge modes to backscatter as long as $[M_{jj'}, S] = 0$. In the case of [1], the set of real matrices commuting with $S$ is spanned by *six* elements, which include the four powers of the symmetry generator $U_y$. The two additional perturbations, rather contrived when written in the $\phi$ basis, *do not* commute with $U_y$ and can be explained by the fact that the zero-frequency components $\zeta_{m,n}^{(0)}, \zeta_{m,n}^{(1)}$ can be coupled without affecting $\zeta_{m,n}^{(2)}, \zeta_{m,n}^{(3)}$. Therefore, one only has $[M_{jj'}, U_y] = 0 \rightarrow [M_{jj'}, S] = 0$ (with the circuit in the main text also satisfying the converse).

## NON-ABELIAN GENERALIZATION

Here we outline the problem of finding the minimal number of nodes per site ($d$) for circuit lattices such that one can simulate unitary hopping matrices of dimension $n$.

Returning to the $n = 2$ and $d = 3$ case, the coupling matrices of the six permutations $(V_y)^\mu$, $P(V_y)^\mu$ (for $\mu = 0, 1, 2$) in the $\phi$ basis form a *permutation representation* of the symmetric group $S_3$, i.e., each matrix has one "1" in each row and column. (In general, the group of permutative connections between two sites, with each site containing $d$ nodes, is the symmetric group $S_d$.) This permutation representation is reducible, i.e., all matrices can be simultaneously block diagonalized into matrices with smaller nonzero blocks. This is exactly what is done by the Fourier transformation into the $\zeta$ basis. In other words, the permutation representation of $S_3$ reduces to a trivial one-dimensional and the desired two-dimensional *ir*reducible *rep*resentation (*irrep*). The one-dimensional part acts on $\zeta^{(0)}$ while the two-dimensional part acts on the pseudo-spin $\langle \zeta^{(1)}, \zeta^{(2)} \rangle$. Upon diagonalization, the onsite capacive coupling $C_0$ isolates the two-dimensional irrep (since $(\tilde{C}_0)_{00} = 0$).

On the other hand, if one uses only nonnegative coefficients (since our variables are inductances), an arbitrary $U(2)$ matrix (in fact, any $2 \times 2$ matrix) can be written as a linear superposition of the 16 elements of a $2 \times 2$ irrep of the Pauli group $\mathcal{P}_1 \approx \{i^\tau I_2, i^\tau \sigma_1, i^\tau \sigma_2, i^\tau \sigma_3\}_{\tau=0}^3$. To realize arbitrary $U(2)$ hoppings, one therefore needs to find the smallest symmetric group $S_d$ that contains a permutation representation of $\mathcal{P}_1$ as its subgroup (formally, the minimal faithful permutation representation, or MFPR, of $\mathcal{P}_1$). This MFPR will then be reduced to the irreps of $\mathcal{P}_1$ via a transformation akin to that from the $\phi$ to the $\zeta$ bases. Finally, the onsite capacitive coupling can be tuned to select the desired irreps. A quick calculation [2] determines that one needs a circuit with $d = 8$ nodes per site in order to construct arbitrary $U(2)$ hopping terms in some $\zeta$ basis.

By a similar procedure, one can realize $U(n > 2)$. To construct arbitrary $n$-dimensional unitary matrices using only real nonnegative coefficients, one first expresses the desired $n \times n$ complex matrix $M$ using the $n^3$ elements of an $n \times n$ irrep of the Generalized Pauli Group $\Pi_n$ [3]. In such an irrep, an element $g_{jkl} = w^j X^k Z^l \in \Pi_n$ where $j, k, l = 0, ..., n-1$; $w = e^{i\frac{2\pi}{n}}$;

$$X = \begin{pmatrix} 0 & 1 & \cdots & 0 \\ \vdots & 0 & 1 & \vdots \\ 0 & & \ddots & 1 \\ 1 & 0 & 0 & 0 \end{pmatrix} \quad \text{and} \quad Z = \begin{pmatrix} 1 & & & \\ & w & & \\ & & \ddots & \\ & & & w^{n-1} \end{pmatrix}.$$

To express $M$ in terms of $g_{jkl}$, notice that $M$ can be expanded in powers of $X$ and $Z$, $M = c_{kl} g_{0kl}$, with complex coefficients $c_{kl} = \frac{1}{n}\text{Tr}\{g_{0kl}^\dagger M\}$. Each $c_{kl}$ can then be expressed as $c_{kl} = d_{jkl} w^j$ with real $d_{jkl} \geq 0$ such that $M = d_{jkl} g_{jkl}$; we note that this decomposition is not unique. Each $g_{jkl}$ corresponds to a specific permutation of inductors and $d_{jkl}$ is the inverse inductance of all of the inductors involved. One then needs to determine the MFPR of $\Pi_n$. This MFPR can then be reduced to irreps of $\Pi_n$ and the capacitive coupling $C_0$ among nodes in a site can be tuned to select the desired irreps, much like $C_0$ had done for the $n = 2$ case above Eq. (3) in the main text. We are fortunate in this case to not have any undesired nontrivial irreps since the irreps of $\Pi_n$ are either variants of the defining $n \times n$ irrep or a trivial $1 \times 1$ irrep [3]. Using an algorithm [4] implemented in Magma [5], we have determined the following dimensions $d$ of the MFPRs of $\Pi_n$.

| $n$ | 3 | 4 | 5 | 6 | 7 | 8 | 9 | 10 | 11 |
|---|---|---|---|---|---|---|---|---|---|
| $d$ | 9 | 16 | 25 | 13 | 49 | 64 | 81 | 29 | 121 |

The evident $n^2$ bound on the MFPR dimension is in fact true for all $n$. This can be seen if one utilizes the proper definition for $d$ [2] to obtain a bound: $d \leq |\Pi_n|/|H|$ for some subgroup $H \subseteq \Pi_n$ satisfying $\text{core}_{\Pi_n}(H) = I = g_{000}$. Here, the core of a subgroup $H$ of $\Pi_n$ is the largest normal subgroup of $\Pi_n$ that is contained in $H$. Our case is simple – we can pick $H = \langle Z \rangle$, the subgroup generated by $Z = g_{001}$. Since $X Z^k X^\dagger \notin \langle Z \rangle$ for $k \neq 0 \mod n$, the core of this subgroup is trivial. Since $|\langle Z \rangle| = n$, this means that $d \leq n^3/n = n^2$.

---


\end{document}